\documentclass{aa}
\usepackage{graphicx}
\usepackage{txfonts}
\usepackage{natbib}
\begin{document}
\newcommand{\be}{\begin{equation}}
\newcommand{\ee}{\end{equation}}
\voffset=-0.25cm

\title{The new pre-cataclysmic
binary PG~2200+085}

\author{
V.~Shimansky\inst{1,2,7},
N.A.~Sakhibullin\inst{1,2,7},
I.~Bikmaev\inst{1,2},
H.~Ritter\inst{3},
V.~Suleimanov\inst{4,1,7},
N.~Borisov\inst{5}, and 
A.~Galeev\inst{1,2,6,7}}

\offprints{V.~Shimansky}
\mail{e-mail: Slava.Shimansky@ksu.ru}

\institute{
Kazan State University, Kremlevskaja str., 18, Kazan 420008, Russia
\and
Tatarstan Academy of Science, Baumana str. 20, 420111, Kazan, Russia
\and
Max-Planck-Institut f\"ur Astrophysik, Karl-Schwarzschild-Strasse 1,D--85740
Garching, Germany
\and
Institut f\"ur Astronomie und Astrophysik, Universit\"at T\"ubingen, Sand 1,
D--72076 T\"ubingen, Germany
\and
Special Astrophysical Observatory, Nizhnij Arkhyz 369167, Russia
\and
Department of Theoretical Physics, Tatar State Humanitarian  Pedagogical
University, Tatarstan str. 2, Kazan 420021, Russia
\and 
Kazan Branch of Isaac Newton Institute, Santiago, Chile
}

\date{Received xxx / Accepted xxx}

   \authorrunning{Shimansky et al.}
   \titlerunning{The new pre-cataclysmic binary PG~2200+085 }

\abstract
{}
{We present the results of spectroscopic-- and orbit--sampled photometric
observations of the faint UV-excess object PG~2200+085.}
{The optical CCD photometry observations of this object were performed
by the Russian-Turkish 1.5-meter telescope RTT150 at the TUBITAK National 
Observatory (Turkey). The long-slit optical spectroscopy observations
with 2.6 \AA ~resolution were carried out by 6-meter telescope BTA at
the Special Astrophysical Observatory (Russia).}
{The
photometric variations over two nights are almost sinusoidal
with an amplitude $\Delta m_V = 0.^{m}04$ and a period of
$P = 0.3186$d. Such a light curve is typical of a detached close
binary with an illumination effect or the ellipsoidal deformation of
a secondary star. The observed spectrum clearly displays a featureless
blue continuum of a hot component and a rich
absorption--line and molecular band K--star spectrum. The CaII line
profiles with strong emission cores are remarkably similar to those of
V471 Tau.}
{We tentatively classify PG~2200+085 as a
pre-cataclysmic binary of the V471 Tau type.}


\maketitle

\section{Introduction}

PG~2200+085 = USNO-A2.0 0975-20754051
($\alpha_{2000}=22^{h}03^{m}19^{s}.72$,
$\delta_{2000}=08^{\circ}45'36''.5$)
is one of the faint ultraviolet sources, $m_V =14.^{m}21$ included in
the Palomar-Green (PG) catalogue  \cite{gr}.  A
finding chart for the object is shown in Fig. 1. The
spectrum of this star was classified as composite, and its colour index
indicates radiation excess in the red spectral region.  Later,
Schultz et al. \cite{schu} found an absorption feature in the spectrum
of PG~2200+085 in the region of the $H_{\alpha}$ line. Apart from
that, PG~2200+085 has not been studied in any further detail.

During the preliminary photometric observations of  PG catalogue objects
that we performed  in 2002,  we found indications of light variations
of PG~2200+085 with an amplitude of $\Delta m_V=0.^{m}02$. Therefore,
in 2003, we performed a more focused  investigation of PG~2200+085,
including longtime photometry with a high-quality CCD--detector
over several nights and spectroscopy with the 6--meter
telescope of the Special Astrophysical Observatory (Russia).

\section[]{Observations}
\subsection{Photometry}

Photometric observations of PG~2200+085 were performed on 
September 18-20 2003, with the  1.5--meter Russian-Turkish
telescope RTT-150 at the TUBITAK National Observatory (Turkey) on
 Bakirlitepe Mountain (see Table 1). The thermoelectrically cooled CCD
camera ANDOR (model DW436, $2048 \times 2048$ with $13.5 \times 13.5~
\mu$m pixels) with an operating temperature of -60$^{\circ}$ C
installed at the Cassegrain focus was used. All observations were
carried out in the V-band with an exposure time of 30 sec and a full
frame readout time of 37 sec (binning 2 x 2). The total observation
time during each night was about 6 hours at an average seeing of 2.5
arcsec. Four stars of similar magnitudes in the field of the CCD ($8'
\times 8'$), which are identified in Fig. 1, have been used as
comparison stars (see Table 2). Five Landolt standard stars have been
observed over both nights, and they have been used to determine the
V-magnitudes of the reference stars in the field around 
PG~2200+085. The brightness of these stars was constant within the
error of the differential photometry of the two objects 
($\Delta m_V=0.^{m}01$).

\begin{table}
\begin{center}
\caption{Log of photometric observations.}
\label{tab1}
\begin{tabular}{cccc}
\hline
Data & Start  & Duration & Number \\
     &  JD    &  hours   & of points \\
\hline
19.09.2003 & 2452901.29470  & 6.25 & 331 \\
20.09.2003 & 2452902.29269  & 6.16 & 327 \\

\hline \end{tabular} \end{center} \end{table}

\begin{table}
\begin{center}
\label{tab1a}
\caption{Variable and comparison stars. Coordinates and magnitudes
are from the USNO-2A catalogue (Monet et al. 1998).
The stars are also identified in Fig.1.}
\begin{tabular}{lccccc}
\hline

Star & V(RTT) &  B & V &  $\alpha_{2000}$ & $\delta_{2000}$ \\
\hline
PG2200 & 14.36 & 14.8 & 14.4 &  22 03 19.7 & +08 45 36  \\
Ref.1 & 14.72 & 14.8 & 14.6 &  22 03 24.4 & +08 47 22 \\
Ref.2  & 14.61 & 15.2 & 14.7 &  22 03 09.2 & +08 45 12 \\
Ref.3  & 14.29 & 14.5 & 14.3 &  22 03 29.3 & +08 43 35 \\
Ref.4  & 14.64 & 15.4 & 14.8 &  22 03 32.6 & +08 46 14 \\
\hline \end{tabular} \end{center} \end{table}

\begin{figure}
\rotatebox{0}{\includegraphics[scale=0.32]{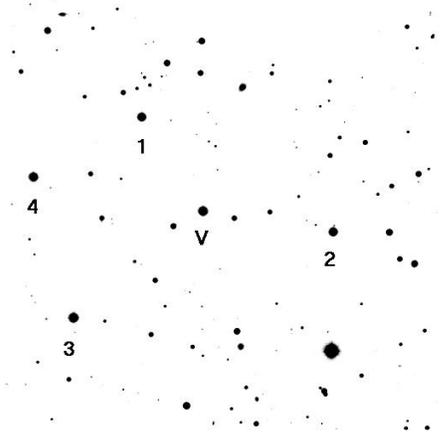}}
\label{fig1a}
\caption{Finding chart of PG~2200+085. North is at the
top  and east to the left. The size is $8' \times 8'$.} \end{figure}

\begin{figure}
\rotatebox{0}{\includegraphics[width=\columnwidth]{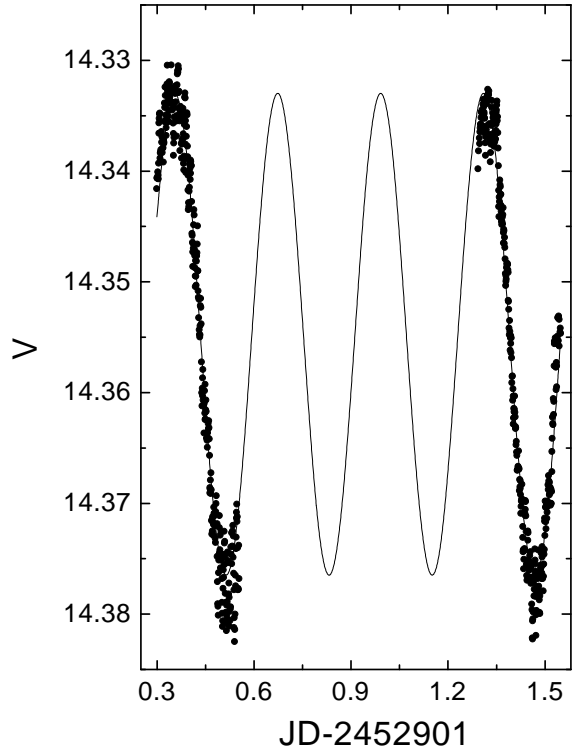}}
\label{fig1}
\caption{The light curve of PG~2200+085 on September 19-20,
2003. The solid curve is the sine function best fitted to the
observed data.}
\end{figure}

\subsection{Spectroscopy}

Spectroscopic observations of PG~2200+085 were carried out at the
6--meter telescope BTA of the Special Astrophysical Observatory with a
long--slit spectrograph \cite{Af} and with a
nitrogen--cooled CCD ($1024 \times 1024$  with $24 \times 24~
\mu$m  pixels) installed at the prime focus. Observations were
performed on August 3, 2003, under excellent weather conditions,
with an average seeing of about 1.$''$3 (see Table 3). We used
a 1302 lines mm$^{-1}$ grating, which gives a  $\Delta\lambda$ = 2.6
\AA~ resolution in the wavelength region  $\Delta \lambda$ 3860--5100
\AA. Three subsequent spectra  with the same exposure time of 600 s
and with a signal-to noise ratio of $S/N \approx 100$ were taken. For
the calibration of wavelengths, fluxes, and radial velocities
$Ar$--$Ne$--$He$ lamp  spectra and spectra of the spectrophotometric
standard star BD +28 2106 from the review by Bohlin \cite{bo} were
taken.

\begin{table}
\begin{center}
\caption{Log of spectroscopic observations.  N: the spectrum number,
 $S/N$ : signal-to-noise ratio, $\varphi_1$ and $\varphi_2$:
orbital phases with respect to ephemeris 1 and 2.}
\label{tab1c}
\begin{tabular}{ccccccc}
\hline
N & UT & HJD & S/N & $\varphi_1$ & $\varphi_2$ \\
\hline
1 & 22$^h$ 16$^m$ & 2452855.4326 & 91 & 0.354 & 0.177 \\
2 & 22$^h$ 26$^m$ & 2452855.4395 & 103 & 0.376 & 0.188 \\
3 & 22$^h$ 37$^m$ & 2452855.4468 & 98 & 0.398 & 0.199 \\
\hline \end{tabular} \end{center} \end{table}

The photometric and spectroscopic data reduction was performed using
the $MIDAS$ \cite{ba} program package.

\begin{figure}
\includegraphics[width=\columnwidth]{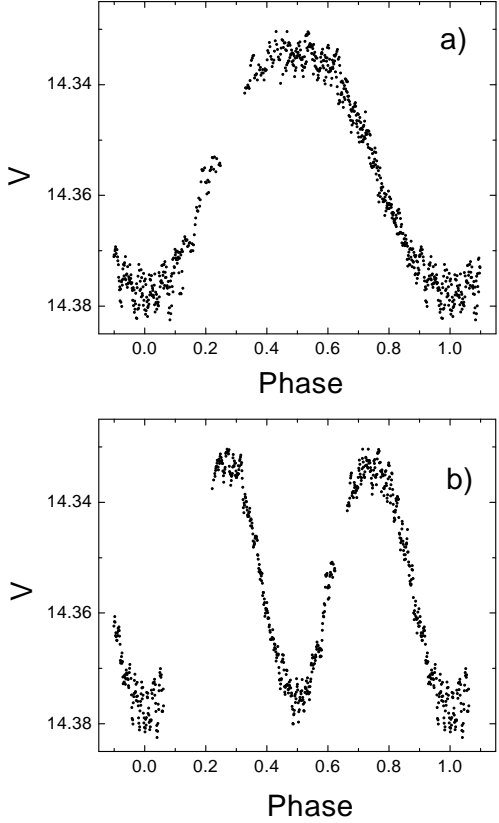}
\caption{\label{fig2}The light curves of PG~2200+085 folded with
the period $0.^{d}31858$ (a) and $0.^{d}63716$ (b).}
\end{figure}

\section[]{Data analysis}

\subsection{The light curves and ephemeris}

The dependence of the magnitude $m_V$ of PG~2200+085 versus the Julian date
of our observations on September 19 and 20, is shown in
Fig. 2. It is obvious that the light curves during
both nights are almost sinusoidal, with a similar amplitude of $\Delta
m_V = 0.^{m}04$.  The fitting of the light curve  with a simple sine  by the $\chi^2$
technique resulted in the following ephemeris:
\begin{equation}\label{equ2}
HJD = 2452900.558 (\pm 0.001) + 0.31858 (\pm 0.00022) \varphi,
\end{equation}
where $\varphi = 0.0$ corresponds to the minimum of $m_V$. The light
curve folded with this period is shown in  Fig. \ref{fig2}a. Such a
light curve is typical of a close binary with a reflection effect
where the ultraviolet radiation of the hot primary component is absorbed
and reemitted in the optical region from the facing
hemisphere of the secondary.

Fitting the light curve by the three--parameter expression
\begin{equation}\label{equ3}
m_V = m_0 + \frac{m_1}{2} * cos(2\pi\varphi) + \frac{m_2}{2} * cos(4\pi\varphi)
\end{equation}
leads to $m_1 = 0.^{m}045 \pm 0.^{m}0003$ and $m_2 << 0.^{m}002$. The
small value of $m_1$ means that either the reflection effect or
the orbital inclination is small, and the very small value of $m_2$
indicates that any superposed ellipsoidal variations in
this light curve are very small.

However, based on the available photometric data, we cannot
exclude the possiblity that the orbital period is twice the value
given in (\ref{equ2}) and that the light curve is double-humped.
The indirect argument in favour of this possibility is the
$10\%$ difference in the amplitude of the variation of
$m_V$ during the nights September 19-20. Correspondingly, the light
curve could also be described (see Fig. \ref{fig2}b) by the ephemeris:
\begin{equation}\label{equ4}
HJD = 2452900.558 (\pm 0.001) + 0.63716 (\pm 0.00051) \varphi.
\end{equation}
Fitting the light curve with Eq. (\ref{equ3}) yields  the
parameters  $m_1 = 0.^{m}004 \pm 0.^{m}0003$ and $m_2
= 0.^{m}044 \pm 0.^{m}0003$. Therefore, if ephemeris
(\ref{equ4}) applies, then  the main light variations ($0.^{m}044$) arise 
due to  ellipsoidal deformation of the secondary.

\begin{figure}
\includegraphics[width=\columnwidth]{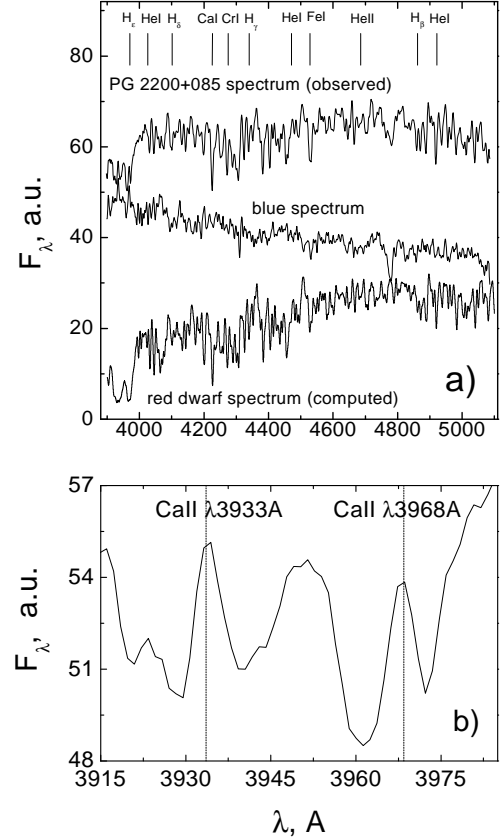}

\caption{\label{fig3}
Observed spectrum of PG~2200+085, theoretical spectrum of the
K--star and blue continuum of the hot component (a);
observed profiles of \rm{CaII$~\lambda \lambda$ 3933, 3968 \AA~} lines (b).}
\end{figure}

\subsection{The spectra}

Cross-correlation analysis of the Doppler line shifts in the
three spectra of PG~2200+085 showed the constancy of the radial
velocities with a measuring accuracy of $\Delta V$ = 3 km s$^{-1}$ during the
30 min of observations. At the same time, all important features in these
spectra had similar intensity.  Therefore, we added the three spectrograms.
This yielded the final spectrum with $S/N > 170$ shown in Fig. \ref{fig3}a.
The presence of strong absorption lines from atoms and ions of heavy
elements and weak molecular bands allows us to classify one of the
components as a main sequence K--star.

To determine the temperature of this star, we computed synthetic
spectra for different values of $T_\mathrm{eff}$ and with
$\log g = 4.5$, $[A] = 0.0$ and compared them to the
observed spectrum. The computations were performed by means of
the program $SPECTR$ \cite{sa}, whereby we
 ignored the possible reflection effect in the
binary. The observed
ratios of the intensities of the metal lines  are well modelled (see
Fig. \ref{fig3}a) with a temperature $T_\mathrm{eff} = 4600 \pm 200K$.
However, the high level of the observed continuum and the low line
intensites indicate the presence of additional flux from the
hot component of PG~2200+085. To extract the contribution of this
component, we subtracted the theoretical spectrum of the K--star
(scaled to the best agreement of observed and theoretical line
intensities) from the observed spectrum of PG~2200+085. The residual
spectrum shown in  Fig. \ref{fig3}a does not show any significant
absorption or emission lines. The strong absorption feature near
$\Delta \lambda$ 4750--4800 \AA~ is the result of neglecting the
molecular bands in the synthetic K--star spectrum. Therefore, this
feature is not part of the spectrum of the hot component. We note that the
real flux distribution of the hot component is possibly even
steeper than shown in  Fig. \ref{fig3}a because we have ignored
interstellar absorption. As a result, we can conclude that the observed
spectrum of PG~2200+085 is composite (as mentioned by Green et al.
(1986), with contributions of a hot white dwarf and a K--star.

At the time of our observations, the radial velocity of
the K--star  was  $V_r = 57 \pm 3 $ km s$^{-1}$. According
to our ephemeris (\ref{equ2}), the spectrograms were taken at
the phases $\varphi$ = 0.34, 0.36 and  0.38. The absence of radial velocity
variations of the K--star between these phases allows us to set an
upper limit on the orbital radial velocity amplitude  of
$V_r$ $\le$ 30 km s$^{-1}$. On the other hand, if we use (\ref{equ4}),
the corresponding  upper limit for the  radial velocity amplitude of
the K--star is 110 km s$^{-1}$.

In the spectrum of PG~2200+085 emission lines of ionized calcium
$\lambda \lambda $ 3933, 3968 \AA~ can be seen ( Fig. \ref{fig3}b).
The measured equivalent widths of these lines are $W_\lambda(3933)$
=1.1 \AA~and $W_\lambda(3968)$=0.9 \AA. These widths are too small for a
typical K--star without flares. Similar intensities of \rm{CaII} emission
lines are, however, seen in published spectra
\cite{sk} of the pre-cataclysmic binary V471 Tau at an orbital
phase $\varphi = 0.25$. Moreover, the intensities of all absorption
lines in the spectra of V471 Tau and PG~2200+085 are very similar.

\section{Interpretation}

 There are a number of possible explanations of the observed basic
features. We begin by recalling them:
\begin{enumerate}
\item The light curve shown in Fig. \ref{fig2} is almost sinusoidal
      with an amplitude of $\Delta V \approx 0.^{m}045$.
\item As can be seen from Fig. \ref{fig3}, even in the restricted
      wavelength range shown the flux from the hot component exceeds
      the flux of the K-star.
\item The system does not eclipse.
\end{enumerate}
In the following we consider three hypotheses, which could explain the
observed light curve.
\begin{enumerate}
\item We see a reflection effect in a close detached binary system
      (a pre-CV). In this case, the orbital period of the system is
      0.3186 d and the reflection effect or the orbital inclination is
      small.
\item We see ellipsoidal variations due to the tidal distortion of the
      K--star in a close detached binary system (a pre-CV). In this
      case, the orbital period of the system is 0.6372 d.
\item We see light variations arising from the rotation of a highly
      magnetized hot white dwarf with an inhomogeneous surface brightness
      distribution, similar to RE J0317-853  \cite{bar1}, in a wide
      binary system. In this case, the period of 0.3186 d reflects the
      white dwarf's rotation period.  
\end{enumerate}
At present, we do not have sufficient observational data to clearly
distinguish between these possibilities. However, there is a statistical
argument against the third hypothesis. Hot ($T_{\rm eff} >$ 40 000 K)
and strongly magnetized white dwarfs, such as RE J0317-853, seem to be
very rare objects, much rarer than pre-CVs. Indeed, only four such
white dwarfs  \cite{wf2000} are currently known.  In addition, none of
them has an unresolved, i.e., very close companion, that can be found by
spectroscopic methods only. From this we may conclude that a priori the
probability for 
PG~2200+085 being a pre-CV is much higher than for a hot and strongly
magnetized white dwarf with a K--star companion. Therefore, we consider
the first two of the above hypotheses more likely. 

Based on the available photometric data alone, and lacking radial
velocity data, it is not possible to distinguish between a reflection
effect and ellipsoidal variations. Here, however, we wish to discuss
arguments that point strongly in favour of the shorter of the two
periods, i.e., that the observed brightness variations are due to a
reflection effect.

 The second observational point allows us to derive a rough lower
limit of the primary's effective temperature. Using Stefan-Boltzmann's
law we get 
\begin{equation}
T_{\rm eff,1} > {\left(\frac{R_2}{R_1}\right)}^{1/2}\, T_{\rm eff,2},
\end{equation}
where $R_1$ and $R_2$ are, respectively, the radius of the primary and the
secondary, and $T_{\rm eff,2}$ the secondary's effective
temperature. Adopting typical values for the white dwarf, i.e.,
$M_1 = 0.6 M_{\odot}$,  $R_1 = 0.01 R_{\odot}$, and
$T_{\rm eff,2} = 4600$K, we obtain a lower limit
$T_{\rm eff,1} > 38500$K. Thus, if the primary is a white dwarf (with
a radius $\sim 0.01R_{\odot}$), then it must be rather hot and,
therefore, could also account for a reflection effect on the facing
hemisphere of the secondary. 
 For our estimate we have assumed that the K--star secondary has a
mass of $0.7 M_{\odot}$ and a radius of $0.7 R_{\odot}$. It is clear
that the resulting primary temperature is only a lower limit. It is
possible to derive an even stronger limit on the temperature of the hot
star. Both stars have comparable fluxes at 5000 - 6000 \AA. At these 
wavelengths, the spectral energy distribution  of the K--star has its
maximum, whereas for the hot star, we are seeing the Rayleigh-Jeans
tail of its spectral energy distribution. Assuming black bodies for
both stars, we can estimate the temperature of the hot component:
\be
   T_1 \approx 6 \cdot 10^5 K~ \left(\frac{\lambda}{6000 \AA}\right)^4
\left(\frac{T_2}{4600 K}\right)^5 \left(\frac{R_2}{0.7R_{\odot}}\right)^2
\left(\frac{R_1}{0.01R_{\odot}}\right)^{-2}
\ee
PG2200+085 is not known to be a soft X-ray source{\footnote 
{http://wave.xray.mpe.mpg.de/rosat/catalogue}},  therefore the hot
component has a temperature less than $\approx 2\cdot 10^5$ K and a
radius larger than $\approx 0.02R_{\odot}$. 

We can also make a simple estimate of the 
amplitude of the resulting reflection effect as follows: Assuming that
the bolometric luminosity of the facing hemisphere of the secondary is
the sum of half the secondary's intrinsic luminosity plus the
illumination luminosity, i.e.,
\begin{equation}
L_{\rm 2,front} = \frac{1}{2} L_{\rm bol,2}
                + \frac{1}{4} {\left(\frac{R_2}{a}\right)}^2\,
                L_{\rm bol,1}
\end{equation}
and that the luminosity of the hemisphere in the shadow of the white
dwarf is
\begin{equation}
L_{\rm 2, back} = \frac{1}{2}\, L_{\rm bol,2},
\end{equation}
we obtain from Eqs. (4)-(6)
\begin{equation}
\frac{L_{\rm 2,front}}{L_{\rm 2,back}} > 1 +
       \frac{1}{2}\, {\left(\frac{R_2}{a}\right)}^2.
\end{equation}
We note that Eq. (7) overestimates the ratio $L_{\rm 2,front}/
L_{\rm 2,back}$ because we have implicitely assumed parallel
illumination rather than illumination from a relatively nearby point
source, and because the secondary loses less than half of its
luminosity over the illuminated hemisphere. In fact, if the
irradiating flux greatly exceeds the secondary's (unperturbed) flux,
energy loss through the irradiated hemisphere will be essentially
suppressed (see, e.g., B\"uning \& Ritter 2004) and the first
term on the right-hand side of Eq. (7) vanishes.

 From our estimate of the K--star parameters, together
with $M_1 = 0.6 M_{\odot}$ and the short orbital period $P = 0.3186$d, 
we obtain $\Delta m_{\rm bol} > 0.057$~mag as a lower limit of the reflection
effect.  This is probably big enough to
account for the observations. If, on the other hand, $P = 0.6372$d,
we still get a contribution from the reflection effect of
$\Delta m_{\rm bol} > 0.02$mag. Such a contribution is, however, not
seen in the observed light curve. Thus, if $P = 0.6372$d, the
modulation, being an almost a pure sine at half that period, must be
entirely due to ellipsoidal variations with essentially no
contribution from a reflection effect. Because the amplitudes of both
effects, i.e., reflection effect and ellipsoidal variations, depend in
a similar way on the orbital incination (both vanish for very low
inclinations), having ellipsoidal varations without a reflection effect
means that there is no reflection effect at all.

Although the above estimate of the amplitude of the reflection effect
shows that there should be some non-negligible contributions, one could
still argue that this estimate is inadequate. Therefore, let us briefly
address the possibility of ellipsoidal variations. Now $P = 0.64716$d.
With the above standard parameters, we find that the radius of the
secondary $R_2$ in units of its Roche radius $R_{\rm 2,R}$ is $\sim
0.52$. Because the tidal deformation of a star that is only about half
the size of its Roche radius is small and because the orbital
inclination can also not be too high (note the absence of eclipses and
the low limit on the radial velocity amplitude), we would argue that
the observed amplitude of the light variations cannot be explained in
terms of ellipsoidal variations.

\begin{figure}

\includegraphics[width=\columnwidth]{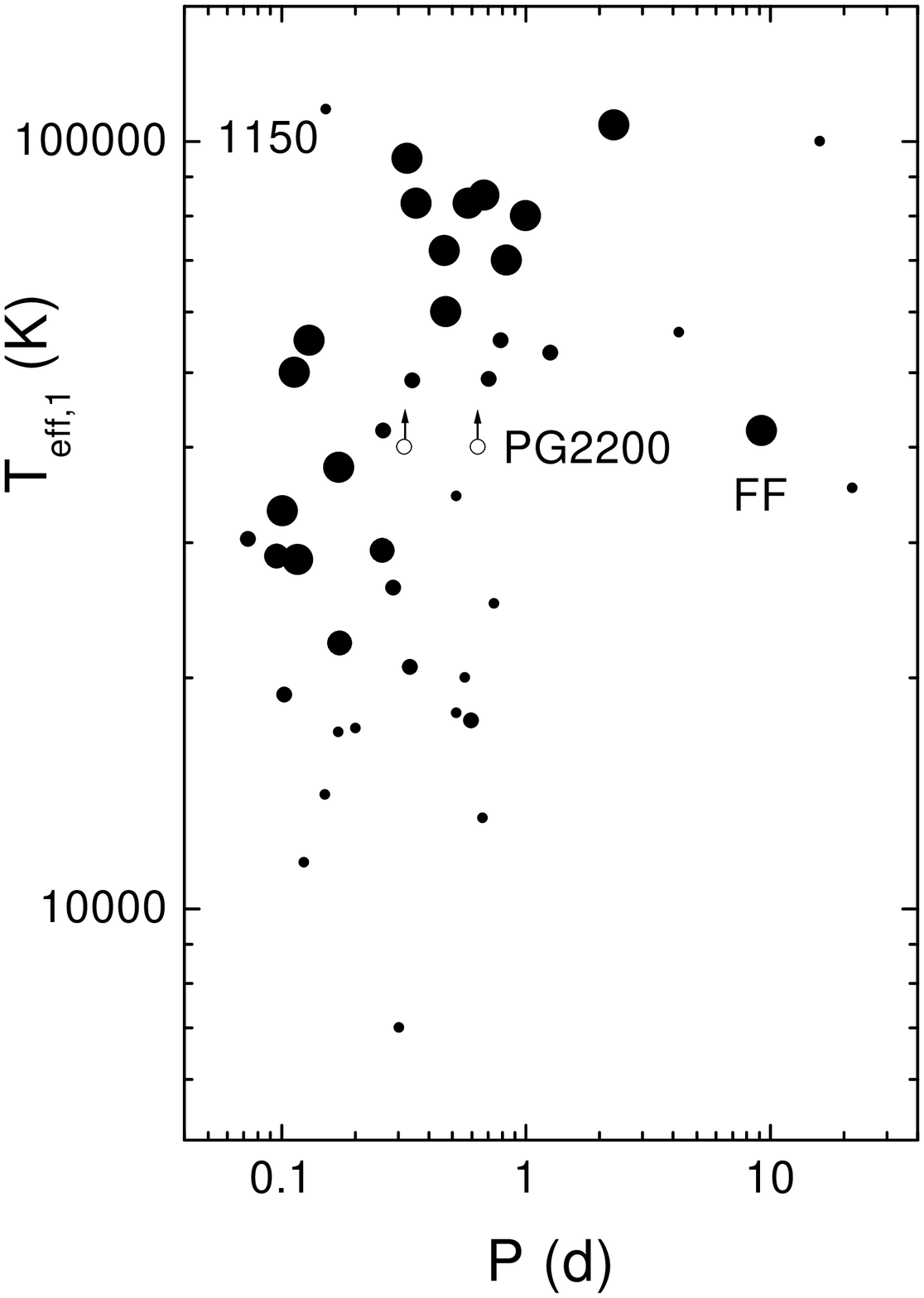}

\caption{Effective temperature $T_{\rm eff,1}$ plotted versus orbital
period for the systems listed in Table \ref{tab4}. The size of the
filled circles indicates the amplitude of the reflection effect. The
smallest symbol is for systems without a reflection effect, the
largest for those where $\Delta m_{\rm refl} > 0.2$mag, the second to
largest for those where $0.2{\rm mag} > \Delta m_{\rm refl} > 0.1
{\rm mag}$, and the second to smallest for those where $0.1{\rm mag} >
\Delta m_{\rm refl} > 0~{\rm mag}$. The lowest position of PG~2200+085
in this diagram is shown by the two open circles. The labels FF and
1150 refer, respectively, to the systems FF~Aqr and 1150+5956.
\label{fig4}}
\end{figure}

Finally, there is an even more compelling argument in favour of the
reflection effect interpretation: In Fig. \ref{fig4} we show the
position of 43 pre-cataclysmic binaries listed in Table \ref{tab4}
in a $T_{\rm eff,1}-P$-diagram, in which the size of each symbol (filled
circles) indicates the amplitude of the observed reflection effect
together with the lowest possible position(s) of PG~2200+085 (open
circles), based on our above estimate of the primary's effective
temperature and the two possible orbital periods. First, we see the
expected trend, namely that the larger the amplitude of the reflection effect,
the higher the effective temperature of the irradiating
star and the lower the orbital period. Second, we see that based on
its position(s) in this diagram, PG~2200+085 is also expected to show
a (weak) reflection effect, even for the longer orbital period (in
agreement with our rough estimate above). The outliers in this diagram
that seem to violate the trend are the systems FF Aqr and 1150+5956.
The contradiction is, however, only apparent. In the case of FF Aqr,
the primary is not only a rather hot sdO star, but it also has a much
larger radius ($\sim 0.16 R_{\odot}$, see Vaccaro \& Wilson
2003) and luminosity than an ordinary white dwarf. This can
fully account for the large reflection effect, even at that long
orbital period, as has been shown by Vaccaro \& Wilson \cite{VW03}.
In the case of 1150+5956, the orbital period is a very preliminary and
detailed time-resolved photometry which could reveal a reflection
effect that is currently unavailable.

\begin{table*}
\begin{minipage}{180mm}
\begin{center}
\caption{Pre-cataclysmic binaries for which the orbital period and the
effective temperature $T_{\rm eff,1}$ of the hot component are
known. A $+$ in the column giving the amplitude of the reflection
effect indicates that a reflection effect of unspecified (and probably
small) amplitude has been seen. }
\label{tab4}
\begin{tabular}{lcllcccl}
\hline
\rule[0mm]{0mm}{5mm}
Object    &Orbital  &Secondary&Primary &$T_{\rm eff,1}$&Reflect.&Filter&References\\
          &period   &spectrum &spectrum&               &effect  &      &          \\

\rule[-3mm]{0mm}{5mm}
          &  (d)    &         &        &($10^3$K)      & (mag.) &      &          \\
\hline
\rule[0mm]{0mm}{5mm}1017-0838 & 0.072994&         &sdB     &$ 30.3 \pm0.1$ & 0.083  &  V   &1         \\
0705+6700 & 0.095647&M~V      &sdB     &$ 28.8 \pm0.9$ & 0.16   &  V   &2         \\
NY Vir    & 0.101016&M~V      &sdB     &$ 33   \pm3  $ & 0.20   &  V   &3         \\
HR Cam    & 0.103063&M~V      &DA3     &$ 19         $ & 0.03   &  V   &4         \\
MT Ser    & 0.113227&         &sdO     &$ 50   \pm5  $ & 0.3    &  B   &5, 6      \\
HW Vir    & 0.116720&         &sdB     &$ 28.5 \pm0.2$ & 0.21   &  V   &7, 8      \\
2237+8154 & 0.123681&M3-4~V   &DA      &$ 11.5 \pm0.5$ &        &      &9         \\
NN Ser    & 0.130080&M6.5~V   &DAO1    &$ 55   \pm8  $ & 0.6    &  V   &10, 11    \\
1347-1258 & 0.150758&M4e~V    &DA      &$ 14.1 \pm0.1$ &        &      &12, 13    \\
1150+5956 & 0.1523: &         &        &$111   \pm10 $&         &      &14, 15    \\
J1129+6637& 0.171   &M4.5~V   &        &$ 17   \pm2  $ &        &      &16        \\
2333+3927 & 0.171802&M3-4~V   &sdB     &$ 37.6 \pm1.0$ & 0.31   &  V   &17        \\
MS Peg    & 0.173666&M3-5~V   &DA2     &$ 22.2 \pm0.1$ & 0.105  &  V   &18, 19    \\
BPM 71214 & 0.201626&M2.5~V   &DA      &$ 17.2 \pm1.0$ &        &      &12        \\
LM Com    & 0.258687&M4+~V    &DA      &$ 29.3       $ & 0.14   &  V   &19, 20    \\
AA Dor    & 0.261582&         &sdO     &$ 42   \pm1  $ & 0.06   &  V   &21, 22    \\
2154+4080 & 0.26772&M~V       &DA2     &$ 30         $ & 0.16   &  V   &23         \\
CC Cet    & 0.286654&M5e~V    &DA2     &$ 26.2 \pm2.0$ & 0.08   &  R   &24, 25    \\
RR Cae    & 0.303700&M5-6~V   &DAwk    &$  7         $ &        &      &26, 27    \\
TW Crv    & 0.32762 &M~V      &sdO     &$105   \pm20 $& 0.846   &  V   &28, 29    \\
1042-6902 & 0.336784&Me~V     &DA3     &$ 20.6 \pm2.0$ & 0.011  &  R   &30, 31    \\
GK Vir    & 0.344331&M3-5~V   &DAO     &$ 48.8 \pm1.2$ &$<$0.05 &  B   &32        \\
KV Vel    & 0.357113&         &sdO     &$ 77   \pm3  $ & 0.55   &  V   &33        \\
UU Sge    & 0.465069&         &sdO     &$ 87   \pm13 $& 0.36    &  V   &34, 35    \\
V477 Lyr  & 0.471729&         &sdO     &$ 60   \pm10 $& 0.62    &  V   &36        \\
J2131+4710& 0.521035625&M4~V  &DA2     &$ 18.0 \pm1.0$ &        &      &37         \\
V471 Tau  & 0.521183&K2~V     &DA2     &$ 34.5 \pm1.0$ &        &      &38        \\
HZ 9      & 0.56433 &M5e~V    &DA2     &$ 17.4       $ &        &      &39, 30    \\
V664 Cas  & 0.581648&G8-K0~V  &sdO     &$ 83   \pm6  $ & 1.15   &  V   &41, 29    \\
UZ Sex    & 0.597259&M4e~V    &DA3     &$ 17.6 \pm2.0$ & 0.012  &  R   &24, 42    \\
EG UMa    & 0.667579&M4-5~V   &DA4     &$ 13.1 \pm0.1$ &        &      &43, 44    \\
VW Pyx    & 0.6758  &         &sdO     &$ 85   \pm6  $ & 1.36   &  V   &45, 46    \\
J2013+4002& 0.70552 &M3-4~V   &DAO     &$ 49.0 \pm0.7$ &  $+$   &      &47, 48    \\
2009+6220 & 0.741226&M5-6~V   &DA2     &$ 25         $ &        &      &31        \\
J1016-0520& 0.78928 &M0-4~V   &DAO     &$ 55   \pm1  $ &  $+$   &      &48, 49    \\
1136+6646 & 0.83607 &K4-7~V   &DAO     &$ 70         $ & 0.24   &  V   &50        \\
Abell 65  & 1.00    &         &sd?     &$ 80         $ &$>$0.5  &  V   &51, 52    \\
IN CMa    & 1.262396&M0-2~V   &DAO     &$ 53   \pm1  $ & 0.075  &  V   &12, 48, 53\\
BE UMa    & 2.291166&K3-4~V   &DAO     &$105   \pm5  $ & 1.3    &  V   &54, 55, 56\\
Feige 24  & 4.23160 &M1-2~V   &DAO     &$ 56   \pm1  $ &        &      &57, 58    \\
FF Aqr    & 9.20803 &G8~III   &sdOB    &$ 42         $ & 0.22   &  U   &58 \\
V651 Mon  &15.991   &A5~V     &        &$100         $ &        &      &60        \\
IK Peg    &21.7217  &A8~V     &DA1     &$ 35.4 \pm0.2$ &        &      &61        \\
\hline \end{tabular} \end{center} 
\end{minipage}

\vspace{0.5cm}

{\it  References:} 1: Maxted et al. \cite{max},  2: Drechsel et.
al. \cite{dre}, 3: Killkenny et al. \cite{kil}, 4:  Maxted et al.
\cite{max1}, 5: Green  et al.  \cite{gre}, 6: Bruch et al.
\cite{bru}, 7: Wood et al. \cite{woo}, 8: Ibanoglu et al.
\cite{iba}, 9: G\"ansicke et al.  \cite{gae}, 10: Catalan et al.
\cite{cat}, 11: Pigulski et al.  \cite{pig}, 12: Kawka et al.
\cite{kaw2}, 13: O'Donoghue et al. \cite{don}, 14: Jacoby et al.
\cite{jac}, 15: Tovmassian et al. \cite{tov}, 16: Raymond et al.
\cite{ray}, 17: Heber et al. \cite{heb}, 18: Schmidt et al.
\cite{sch}, 19: Shimansky et al. \cite{shi}, 20: Orosz et al.
\cite{oro}, 21: Rauch  \cite{rau}, 22: Hilditch et al.
\cite{hil}, 23: Hillwig et al. \cite{hil2}, 24: Saffer et al.
\cite{saf}, 25: Somers et al. \cite{som}, 26: Bragaglia et al.
\cite{bra}, 27: Bruch et al. \cite{bru}, 28: Chen et al.
\cite{che}, 29: Exter et al. \cite{ext}, 30: Kawka et al.
\cite{kaw1}, 31: Morales-Rueda et al. \cite{max3}, 32: Fulbright et al.
\cite{ful}, 33: Hilditch et al. \cite{hil}, 34: Pollacco et al.
\cite{pol}, 35: Bell et al. \cite{bel}, 36: Pollacco et al.
\cite{pol1}, 37: Maxted et al. \cite{max2} 38: O'Brien et al.
\cite{bri}, 39: Lanning et al. \cite{lan}, 40: Schreiber et al.
\cite{sg}, 41: Shimanskii et al.  \cite{shi1}, 42: Bruch et al.
\cite{bru1}, 43: Bleach et al.  \cite{ble}, 44: Bleach et al.
\cite{ble1}, 45: Kohoutek et al. \cite{koh}, 46: Exter et al.
\cite{ext}, 47: Thorstensen et al. \cite{tho}, 48: Vennes et al.
\cite{ven}, 49: Thorstensen et al. \cite{tho1}, 50: Sing et al.
\cite{sin}, 51: Bond et al. \cite{bon}, 52: Walsh et al.
\cite{wal}, 53: Barstow et al. \cite{bar}, 54: Wood et al.
\cite{woo1}, 55: Ferguson et al. \cite{fer}, 56: Raguzova et al.
\cite{rag}, 57: Vennes et al. \cite{ven1}, 58: Vennes et al.
\cite{ven2}, 59: Vaccaro et al. \cite{VW03}, 60: Mendez et al.
\cite{men}, 61: Vennes et al. \cite{ven3}.

\end{table*}

From these arguments, we would then conlude that the true orbital period
of PG~2200+085 is $P = 0.3186$d and that the observed periodic
modulation is the result of a reflection effect on the K--star caused
by the hot primary white dwarf. The comparatively small amplitude of
the reflection effect, in turn, results from the relatively low
inclination of the system.

\section{Conclusions}

The observations presented in this paper allow us to conclude that
PG~2200+085 is a pre-cataclysmic binary consisting of a hot (
pre-) white dwarf and a K--dwarf companion, and with an orbital
period of either $P = 0.3186$d or $P= 0.6372$d. When comparing
PG~2200+085 to the pre-cataclysmic binaries listed by Schreiber
\& G\"ansicke \cite{sg}, or to the most recent version (release 7.4)
of the Ritter \& Kolb \cite{ri} catalogue, we note that PG~2200+085 is
only the second such system consisting of a white dwarf and a K--star
companion (the other one being V471~Tau). All other known short--period
pre-cataclysmic binaries either consist of a white dwarf and an
M--dwarf companion, or, if the companion is a K--star, the hot
component is an extremely hot sdO star. From the evolutionary point of
view, systems like V471~Tau and PG~2200+085 are intersting because
they are the direct progenitors of cataclysmic variables with a
comparatively long orbital period, i.e., $P \ge 6$hr. 
 We cannot exclude the other possibility, namely, that this object
is a wide binary consisting of a hot, highly magnetized white dwarf
and a K--star companion. But, according to the statistical arguments
presented above, this is much less probable. Nonetheless, if it were true,
PG~2200+085 is an even more interesting object than a pre-CV.
For this reason alone, PG~2200+085 deserves further and detailed study. 

 In the framework of the pre-CV hypothesis, we have presented
theoretical and observational arguments that
resolve the orbital period ambiguity in favour of the shorter period
$P = 0.3186$d, i.e., we favour the interpretation that the optical
modulation is due to a reflection effect on the facing hemisphere of
the K--star, which, in turn, is caused by irradiation from the
moderately hot ($T_{\rm eff,1} \ge 38500$ K) white dwarf. However, for
a final decision about which of the two proposed periods is the correct
one, a detailed radial velocity study of PG~2200+085 is required.

\section*{Acknowledgements}

The authors are grateful to the Large Telescopes Program Committee
for many years of support of our observational programs, and to
the referee, Dr. Barstow, for his interesting suggestion that
PG~2200+085 could be a hot and highly magnetized white dwarf.
This work was supported by the Russian Foundation of Fundamental
Research (grant $05$--$02$--$17744$) and by the President program for
support of the leading science school (grant NSh - 784.2006.2). VS
thanks DFG for partial financial support (grant We 1312/35-1).

\label{lastpage}

\end{document}